\documentclass{proc}
\usepackage{epsfig,float}
\usepackage[l]{floatflt}
\renewcommand{\bottomfraction}{0.99}

\onecolumn

\def\to{\rightarrow}

\def\te{\tilde e}

\def\tb{\tilde b}

\def\tst{\tilde t}
\def\ttau{\tilde \tau}
\def\tmu{\tilde \mu}

\def\tw{\widetilde W}
\def\tz{\widetilde Z}
\def\alt{\stackrel{<}{\sim}}

\begin{document}
\renewcommand{\textfraction}{0.0}
\renewcommand{\bottomfraction}{1.0}

\begin{center}
{\Large\bf Relic density of neutralinos in minimal supergravity 
\footnote{
Talk given by Alexander Belyaev at SUSY'02, 
"The 10th International Conference
on Supersymmetry and Unification of Fundamental Interactions", 
DESY, Hamburg, Germany, 17-23 June 2002}
\\
\ \ }
\vskip 0.3cm
{Howard Baer, Csaba Bal\'azs and Alexander Belyaev\footnote
{On leave of absence from Nuclear Physics Institute, Moscow State University.}\\
\vskip 0.2cm
	{\it Department of Physics, Florida State University, 
                  Tallahassee, FL, USA 32306}\\
	E-mail: baer@hep.fsu.edu, balazs@hep.fsu.edu, belyaev@hep.fsu.edu}

\end{center}

\abstract{We evaluate the relic density of neutralinos in the minimal
supergravity (mSUGRA) model.  
All $2\rightarrow 2$ 
neutralino annihilation diagrams, as well as all 
initial states involving sleptons, charginos, neutralinos and third generation squarks are included.
Relativistic thermal averaging of the velocity times cross sections is performed.
We find that co-annihilation effects are only important on the edges of the model
parameter space, where some amount of fine-tuning is necessary to obtain a 
reasonable relic density. Alternatively, at high $\tan\beta$, 
annihilation through the broad Higgs resonances gives rise to
an acceptable neutralino relic density over broad regions of parameter
space where little or no fine-tuning is needed.}

\section{Introduction}

A wide variety of astrophysical measurements are being used to pin down some of the basic
cosmological parameters of the universe. High resolution maps of the cosmic microwave background
(CMB)  radiation\cite{cmb} 
imply that the energy density of the universe $\Omega=\rho/\rho_c \simeq 1$,  consistent with
inflationary cosmology. Here, $\rho_c= 3 H^2/8\pi G_N$ is the critical closure density of the
universe, where $G_N$ is Newton's constant and $H=100h$ km/sec/Mpc is the scaled Hubble constant. The
value of $h$ itself is determined to be $\sim 0.7\pm 0.1$ by improved measurements of distant
galaxies\cite{wlf}.  Meanwhile, data from distant  supernovae\cite{supernovae} imply a nonzero dark
energy content of the universe $\Omega_\Lambda \sim 0.7$, a result which is confirmed  by fits to the
CMB power spectrum\cite{jaffe}. Analyses of Big Bang nucleosynthesis\cite{nucleo} imply the baryonic
density  
$\Omega_b h^2\simeq 0.020\pm 0.002$, 
although the CMB fits suggest a  somewhat higher value of $\sim 0.03$. Hot
dark matter, for instance  from massive neutrinos, should give only a small contribution to the total
matter density of the universe. In contrast, a variety of data ranging from galactic rotation curves
to  large scale structure and the CMB imply a significant density of cold dark matter
(CDM)\cite{review} $\Omega_c h^2\simeq 0.2\pm 0.1$.

In many $R$-parity conserving supersymmetric models of particle physics, the lightest neutralino
($\tz_1$)  is also the lightest SUSY particle (LSP); as such, it is massive, neutral and stable. For
this case, relic neutralinos left over from the Big Bang provide an  excellent candidate for the CDM
content of the universe\cite{jkg}.
In this work, we  present results of calculations of the neutralino relic density within the
context of the paradigm minimal supergravity model (mSUGRA, or CMSSM)\cite{sugra}. In mSUGRA, it
is assumed that SUSY breaking occurs in a hidden sector of the model,  with SUSY breaking
effects communicated from hidden to observable sectors via  gravitational interactions. The
model parameter space is given by %
\begin{equation}
m_0,\ m_{1/2},\ A_0,\ \tan\beta\ {\rm and}\ sign(\mu ) .
\end{equation}
Here, $m_0$ is the universal scalar mass,  $m_{1/2}$ is the universal gaugino mass and $A_0$ is
the universal trilinear mass all evaluated at $M_{GUT}$, while $\tan\beta$ is the ratio of Higgs
field vevs ($v_u/v_d$), and $\mu$ is a supersymmetric Higgs mass term. The soft SUSY breaking
parameters, along with gauge and Yukawa couplings, evolve from $M_{GUT}$ to $M_{weak}$ according
to their renormalization group (RG) equations. At $M_{weak}$, 
the RG improved 1-loop effective potential is minimized, and
electroweak gauge symmetry is broken radiatively. In this report, we  implement the mSUGRA
solution encoded in ISAJET v7.64\cite{isajet}.

There is a long history of increasingly sophisticated solutions for the relic density of
neutralinos in supersymmetric  models\cite{early,ows,barb,gkt,gs,gg,bottino,dn,%
leszek,an,bb,eg,bk,ellis,darksusy,fmw,ellis_co,an2,pallis, leszek2,manuel,drees,santoso,belanger,roszkowski}.  
The
key ingredient to  solving the Boltzmann equation is to evaluate the thermally averaged
neutralino annihilation cross section times velocity factor. Traditionally, the solution is made
by expanding the annihilation  cross section as a  power series in neutralino velocity,  so that
angular and energy integrals can be evaluated analytically. The remaining integral over
temperature can then be performed numerically. The power series solution is valid  in many
regions of model parameter space because the relic neutralino  velocity is expected to be highly
non-relativistic.

However, it was emphasized by Griest and Seckel that annihilations may occur through $s$-channel
resonances at high enough energies\cite{gs} that a relativistic treatment of thermal averaging might
be necessary.  Drees and Nojiri found that at large values of the parameter $\tan\beta$,  neutralino
annihilation can be dominated by $s$-channel scattering through  broad $A$ and $H$ Higgs
resonances\cite{dn}.  The proper formalism for relativistic thermal averaging was developed by
Gondolo and Gelmini (GG)\cite{gg},  and was implemented in the code of Baer and Brhlik\cite{bb,bk}. 
Working within the framework of the mSUGRA model,  it was found\cite{bb,bk,ellis,leszek2,manuel}
that at large $\tan\beta$,  indeed large new regions of model parameter space gave rise to
reasonable values for the CDM relic density.  At large $\tan\beta$, the $A$ and $H$ resonances are
broad enough (typically 10-50 GeV) that even if  the quantity $2m_{\tz_1}$ is several partial widths
away from exact resonance, there can still be a significant rate for neutralino annihilation.  Thus,
in the mSUGRA model at low $m_0$ and $\tan\beta$,  neutralino annihilation is dominated by
$t$-channel slepton exchange, and reasonable values of the relic density occur only for relatively
low values of $m_0$ and $m_{1/2}$. At high $\tan\beta$, a much larger parameter space is allowed,
owing to off-resonance neutralino annihilation through the broad Higgs resonances.

In addition, there exist regions of mSUGRA model parameter space where co-an\-ni\-hi\-la\-tion
processes are important, and even dominant.  It was stressed by Griest and Seckel\cite{gs} that in
regions with a higgsino-like LSP,  the $\tz_1$, $\tw_1$ and $\tz_2$ masses become nearly degenerate,
so that all three species can exist in thermal equilibrium, and annihilate against  one another. The
relativistic thermal averaging formalism of GG was  extended to include co-annihilation processes by
Edsj\"o and Gondolo\cite{eg},  and was implemented in the DarkSUSY code\cite{darksusy}  for
co-annihilation of charginos and heavier neutralinos.

The importance of neutralino-slepton co-annihilation was stressed by Ellis {\it et al.} and
others\cite{ellis_co,an2,pallis,leszek2,manuel}.  In regions of mSUGRA parameter space
where $\tz_1$ and $\ttau_1$ (or other sleptons) were nearly  degenerate (at low $m_0$),
co-annihilations could give rise to reasonable values of the relic density even at very large values
of $m_{1/2}$, at both low and high $\tan\beta$. In addition, for large values of the parameter $A_0$
or for non-universal scalar masses, top or bottom squark masses could become nearly degenerate  with
the $\tz_1$, so that squark co-annihilation processes can become important as
well\cite{drees,santoso}.

The goal of this study is to calculate the relic density of  neutralinos in the mSUGRA model including
co-annihilation processes in addition to {\it relativistic} thermal averaging of  the annihilation
cross section times velocity.  Since there are very many Feynman diagrams to evaluate for neutralino
annihilations and co-annihilations, we use  CompHEP~v.33.23\cite{comphep}, which provides for fast
and efficient  automatic evaluation of tree level processes in the SM or MSSM. For initial states
including $\tz_1$, $\tz_2$, $\tw_1$, $\te_1$, $\tmu_1$, $\ttau_1$, $\tst_1$ and $\tb_1$, we count
1722 subprocesses,  including 7618 Feynman diagrams. For those processes we have  calculated the
squared matrix element and have written it down  in the form of CompHEP {\it FORTRAN} output.

The weak scale parameters from supersymmetric models are generated using ISAJET v7.64,  and interfaced
with the squared matrix elements from CompHEP. Details of our computational algorithm are given in Sec.
2. In Sec. 3 we present a variety of results for the relic density in mSUGRA model parameter space.
Much of parameter space is ruled out at low $\tan\beta$ since the relic density is too high, and would
yield too small an age of the universe. At high $\tan\beta$, large regions of parameter space are
available with a reasonable relic density in the range $0.1< \Omega_{\tz_1} h^2 < 0.3$.
In Sec. 4, we conclude.
\vskip -0.5cm
\section{Calculational Details}
The evolution of the number density of supersymmetric relics in the
universe is described by the Boltzmann equation
as formulated for a 
Friedmann-Robertson-Walker universe. For calculations including 
many particle species, such as the case where co-annihilations
are important, there is a Boltzmann equation for each particle species.
Following Griest and Seckel\cite{gs}, the equations can be combined to obtain
a single equation
\begin{equation}
\frac{dn}{dt} =-3Hn-\langle\sigma_{eff}v\rangle\left(n^2-n_{eq}^2\right),\ 
\mbox{where \ \ }
n=\sum_{i=1}^{N} n_i
\end{equation}
and the sum extends over the $N$ particle species contributing to
the relic density, with $n_i$ being the number density of the $i$th
species.
Furthermore, $n_{eq,i}$ is the number density of the $i$th species 
in thermal equilibrium, given by
\begin{equation}
n_{eq,i} =\frac{g_im_i^2T}{2\pi^2}K_2\left(\frac{m_i}{T}\right) ,
\end{equation}
where $K_j$ is a modified Bessel function of the second kind of order $j$.

The quantity $\langle\sigma_{eff}v\rangle$ is the thermally averaged
cross section times velocity. A succinct expression for this quantity
using relativistic thermal averaging was computed by Gondolo and Gelmini
for the case of a single particle species\cite{gg}, and was extended by
Edsj\"o and Gondolo for the case including co-annihilations\cite{eg}.
We adopt this latter form, given by
\begin{equation}
\langle\sigma_{eff}v\rangle (x) =\frac{\int_2^\infty K_1\left({a\over x}
\right)
\sum_{i,j=1}^{N}\lambda(a^2,b_i^2,b_j^2)g_ig_j\sigma_{ij}(a)da}
{4x\left( \sum_{i=1}^{N}K_2\left({b_i\over x}\right) b_i^2 g_i\right)^2} ,
\end{equation}
where $x=T/m_{\tz_1}$ is the temperature in units of mass of the relic 
neutralino, $\sigma_{ij}$ is the cross section for the annihilation
reaction $ij\to X$ ($X$ is any allowed final state consisting of 
2 SM and/or Higgs particles), 
$\lambda(a^2,b_i^2,b_j^2) = 
a^4+b_i^4+b_j^4-2(a^2 b_i^2+a^2 b_j^2+ b_i^2 b_j^2)$, 
$a=\sqrt{s}/m_{\tz_1}$ and $b_i=m_i/m_{\tz_1}$.
This expression is our master formula for the relativistically
thermal averaged annihilation cross section times velocity.

To solve the Boltzmann equation, we introduce a freeze-out
temperature $T_F$, so that the relic density of neutralinos is given 
by\footnote{The procedure we follow gives numerical results valid
to about 10\% versus a direct numerical solution of the 
Boltzmann equation\cite{gg}.}
\begin{equation}
\Omega_{\tz_1} h^2= \frac{\rho (T_0)}{8.1\times 10^{-47}\ {\rm GeV}^4}
\end{equation}
where
\begin{equation}
\rho (T_0)\simeq 1.66 {1\over M_{Pl}}
\left(\frac{T_{m_{\tz_1}}}{T_\gamma}\right)^3 T_\gamma^3\sqrt{g_*}
\frac{1}{\int_0^{x_F}\langle \sigma_{eff}v\rangle dx} .
\end{equation}
The freeze-out temperature $x_F=T_F/m_{\tz_1}$ 
is determined as usual by an iterative
solution of the freeze-out relation
\begin{equation}
x_F^{-1}=\log\left[ \frac{m_{\tz_1}}{2\pi^3}\frac{g_{eff}}{2}
\sqrt{\frac{45}{2g_* G_N}}\langle\sigma_{eff}v\rangle (x_F) \, x_F^{1/2} \right] .
\end{equation}
Here, $g_{eff}$ denotes the effective number of degrees of freedom
of the co-annihilating particles, as defined by Griest and Seckel\cite{gs}.
The quantity $g_*$ is the SM effective degrees of freedom parameter
with $\sqrt{g_*}\simeq 9$ over our region of interest.

The challenge then is to evaluate all possible channels for
neutralino annihilation  to SM and/or Higgs particles, as well as
all co-annihilation reactions. The 7618 Feynman diagrams are evaluated
using CompHEP.
 To achieve our final result with relativistic thermal
averaging, a three-dimensional integral must be performed over
{\it i.}) the final state subprocess scattering angle $\theta$, 
{\it ii.}) the subprocess energy parameter $a=\sqrt{s}/m_{\tz_1}$,
and {\it iii.}) the temperature $T$ from freeze-out $T_F$ to the present
day temperature of the universe, which can effectively be taken to be 0.
We perform the three-dimensional integral using the BASES 
algorithm\cite{bases},
which implements sequentially improved sampling in multi-dimensional
Monte Carlo integration, generally with good convergence
properties.
We note that the three-dimensional integration
appearing in the case of our relativistic calculations involving
several species in thermal equilibrium is about 2 orders of magnitude more 
CPU-time consuming than the series expansion approach, 
which requires just one numerical integration.
\section{Results}
\begin{figure}[h]
\vskip -0.5cm
\epsfig{file=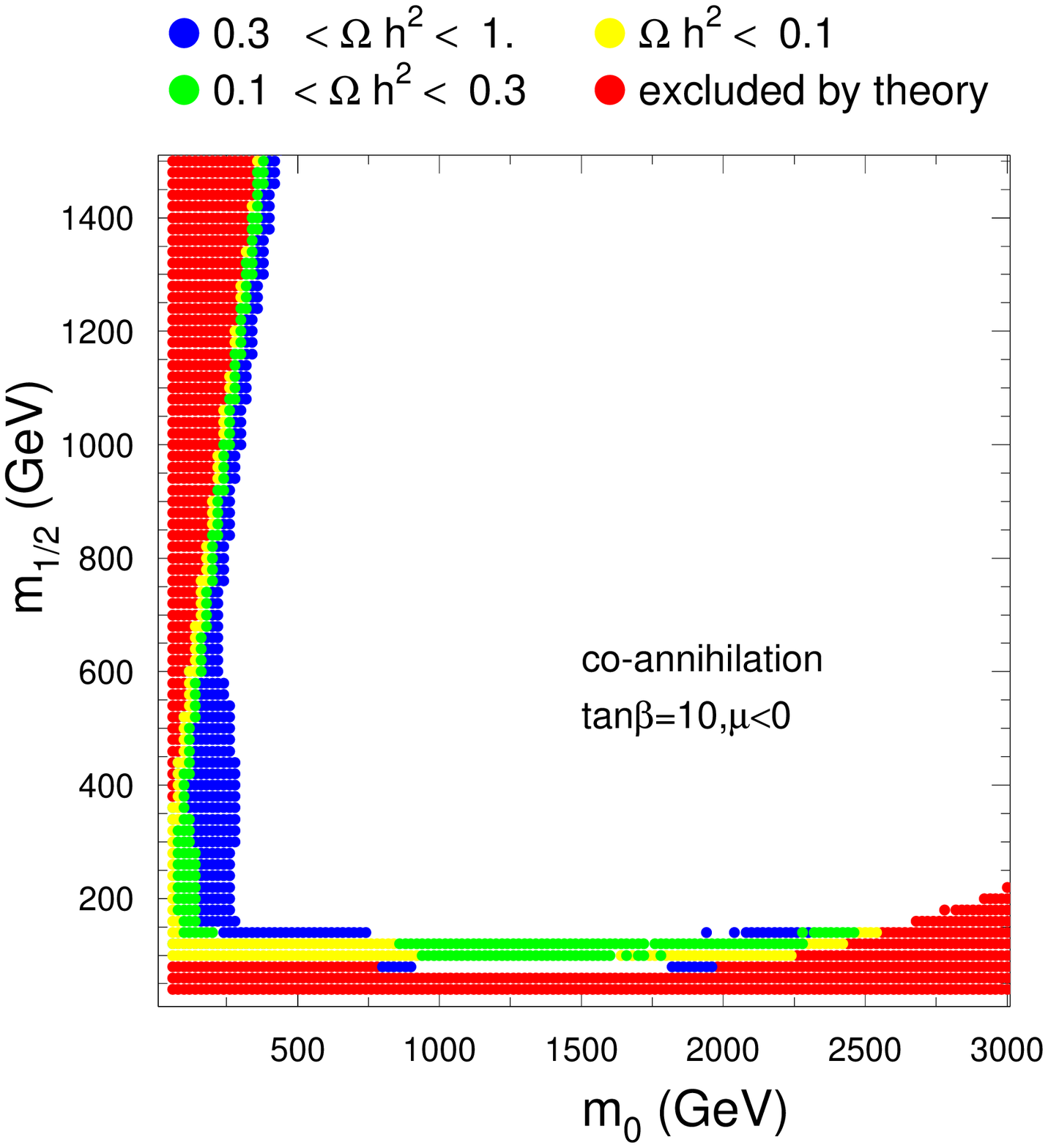,width=8cm}
\hspace*{-0.5cm}
\epsfig{file=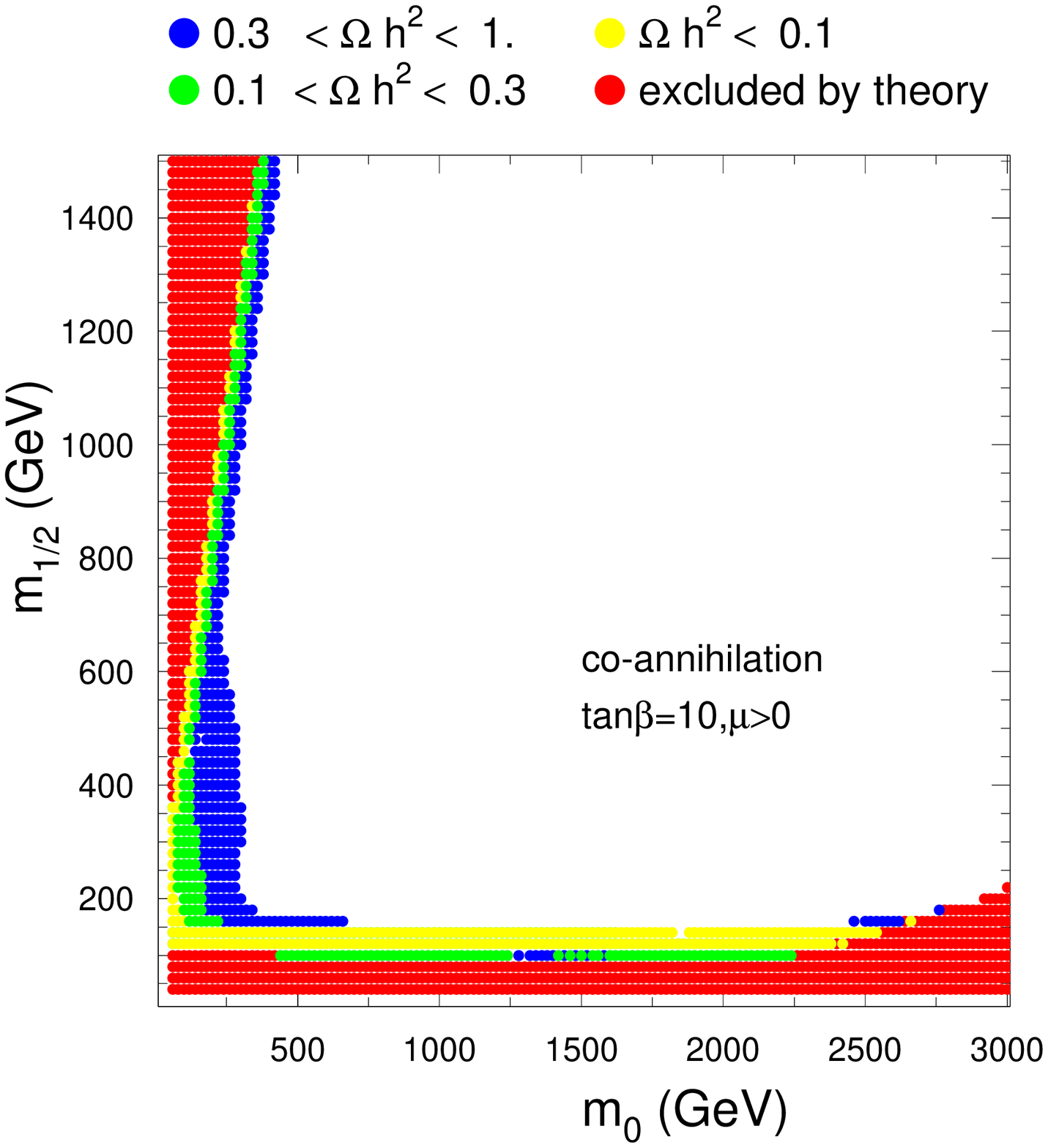,width=8cm}
\vskip -0.7cm
\caption{Regions of neutralino relic density in the 
$m_0\ vs.\ m_{1/2}$ plane for $A_0=0$ and $\tan\beta =10$.}
\label{plane_10}
\vskip -0.3cm
\end{figure}

\begin{figure}[h]
\vskip -0.5cm
\epsfxsize=8cm \epsfbox{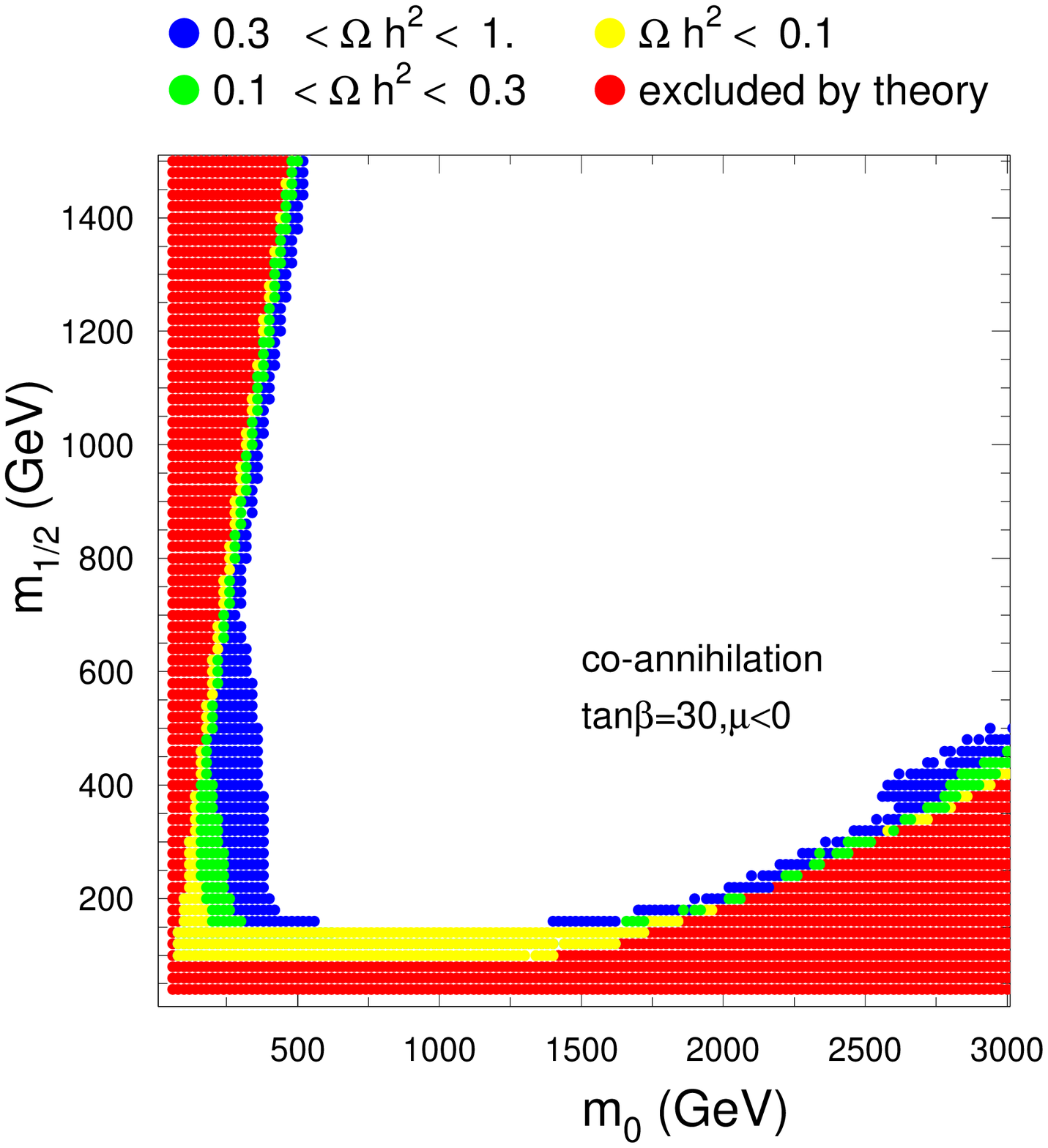}
\epsfxsize=8cm \epsfbox{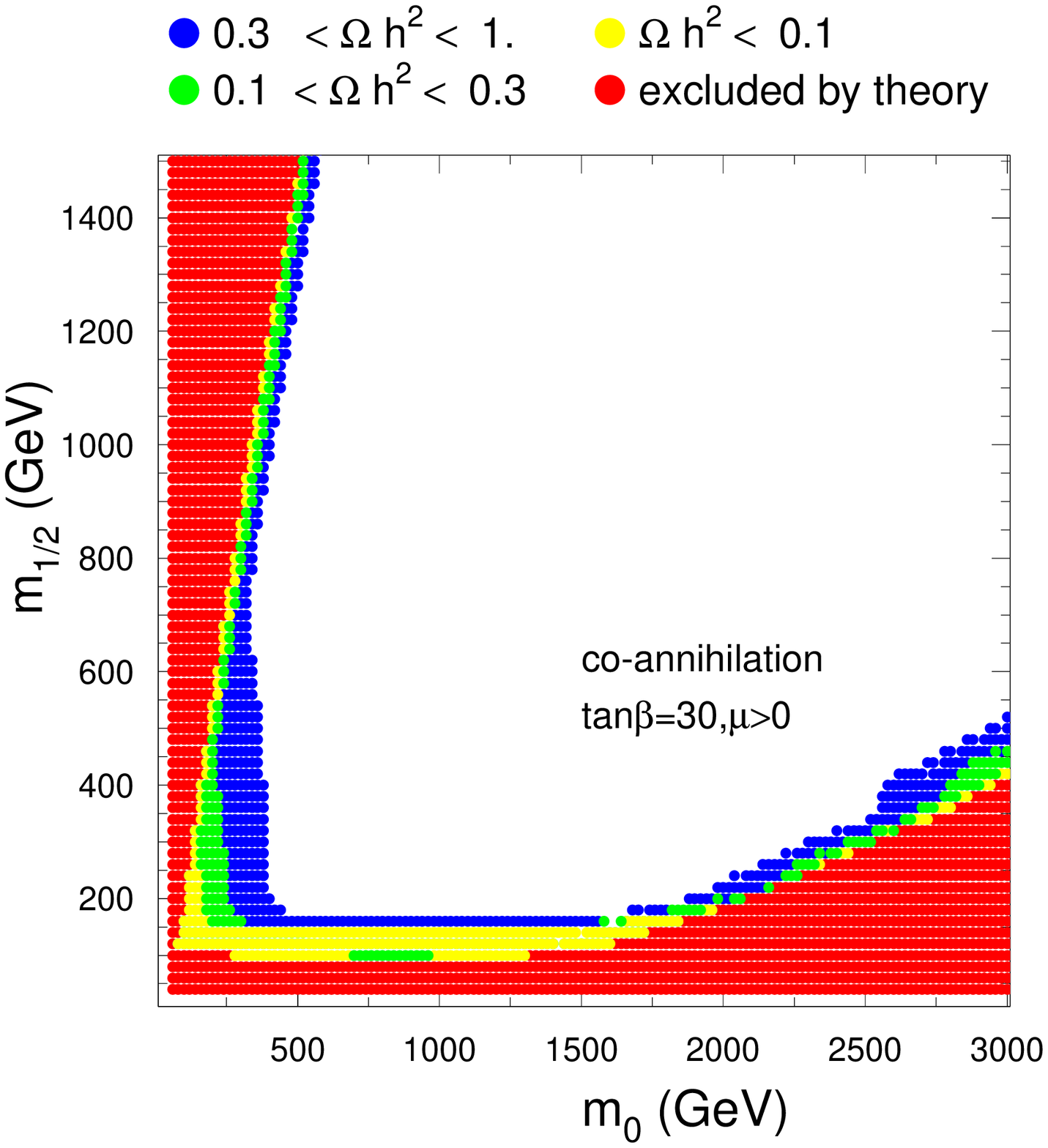}
\vskip -0.7cm
 \caption{Regions of neutralino relic density in the 
  $m_0\ vs.\ m_{1/2}$ plane for $A_0=0$ and $\tan\beta =30$.}
  \label{plane_30}
\vskip -0.1cm
\end{figure}

Our first results in Fig.~\ref{plane_10} show regions of 
$\Omega_{\tz_1} h^2$ in the 
$m_0\ vs.\ m_{1/2}$ plane in the minimal supergravity model
for $A_0=0$, $\tan\beta =10$ and for $\mu<0$(left) and $\mu >0$(right).
\begin{floatingfigure}{7.7cm}
\vskip -0.5cm
\hspace*{-0.7cm}\epsfig{file=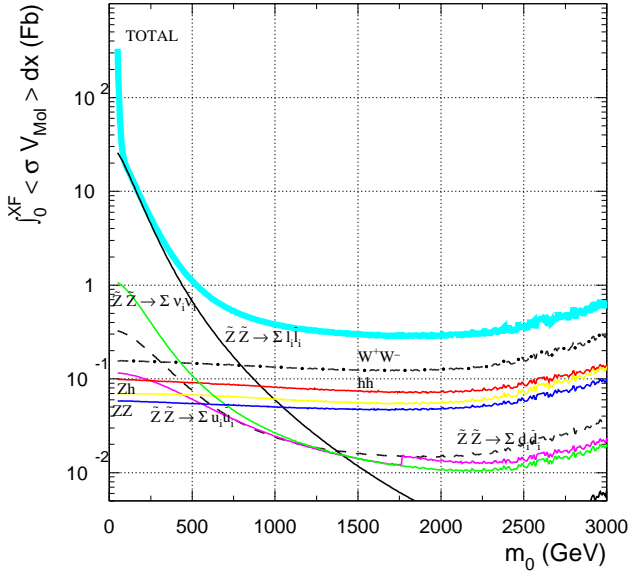,width=8.5cm}
\vskip -0.7cm
\caption{Thermally averaged cross section times 
velocity integrated from $T=0$ to $T_F$, for various 
subprocess. The thick light-grey(light-blue) curve denotes the
total of all annihilation and co-annihilation
reactions; $m_{1/2}=300$ GeV, $\mu >0$, $A_0=0$ 
and $\tan\beta =10$.}%
\label{1d_sp300_10}
\end{floatingfigure}
The dark shaded (red) regions are excluded by theoretical
constraints (lack of REWSB on the right, a charged LSP in the upper left).
The unshaded regions have $\Omega_{\tz_1}h^2 >1$, and should be excluded, 
as they would lead to a universe of age less
than 10 billion years, in conflict with the oldest stars found in globular
clusters. 
The medium shaded (green) region
yields values of $0.1< \Omega_{\tz_1} h^2 <0.3$, {\it i.e.} in the most 
cosmologically favored region. The  light shaded (yellow)($\Omega_{\tz_1} h^2 <0.1$) and
black(blue) ($0.3< \Omega_{\tz_1} h^2 <1$) correspond to regions with 
intermediate values of low and high relic density, respectively. 
Points with $m_{1/2}\alt 150$ GeV give rise to chargino masses below
bounds from LEP2; the LEP2 excluded regions due to chargino, slepton and Higgs
searches are not shown on these plots.
The structure of these plots can be understood by examining
the thermally averaged cross section times velocity, integrated from
zero temperature to $T_F$. 
In Fig.~\ref{1d_sp300_10} we show this quantity
for a variety of contributing subprocesses plotted versus $m_0$
for fixed $m_{1/2}=300$ GeV, $\mu >0$, and all other parameters as in 
Fig.~\ref{plane_10}. At low values of $m_0$, the neutralino annihilation
cross section is dominated by $t$-channel scattering into leptons pairs,
as shown by the black solid curve. 
However, at the very lowest values of $m_0$,
the annihilation rate is sharply increased by neutralino-stau and stau-stau
co-annihilations, leading to very low relic densities where 
$m_{\tz_1}\simeq m_{\ttau_1}$\cite{ellis_co}.
 As $m_0$ increases, the
slepton masses also increase, which suppresses the annihilation
cross section, and the relic density rises to values $\Omega_{\tz_1}h^2>1$.
When $m_0$ increases further, to beyond the $\sim 1$ TeV level, 
and approaches the excluded region, the
magnitude of the $\mu$ parameter falls, and the higgsino component of
$\tz_1$ increases.
This is the so called ``focus point'' region, explored in
Ref. \cite{fmw}.
In this region, the annihilation rate
is dominated by scattering into $WW$, $ZZ$, $hh$  and $Zh$ channels. 
At even higher $m_0$ values, $m_{\tz_1}\simeq m_{\tw_1}\simeq m_{\tz_2}$,
and these co-annihilation channels increase even more the annihilation rate.
Finally, at the large $m_0$ bound on parameter space, $|\mu |\to 0$,
and appropriate REWSB no longer occurs.
Most of the structure of Fig.~\ref{plane_10} can be understood in these
terms, with the exception being the horizontal band of very low
relic density at $m_{1/2}\simeq 125$ GeV. 
In this region, which is nearly
excluded by LEP2 bounds on the chargino mass, there is enhanced 
neutralino annihilation through the $Z$ and $h$ resonances. In fact, 
a higher degree of resolution on our plots would resolve 
these horizontal bands into {\it two} bands, corresponding to each of the 
separate resonances, as shown in Ref. \cite{bb}.
The $m_0\ vs.\ m_{1/2}$ planes for $\tan\beta =30$ are shown 
in Fig.\ref{plane_30}. The structure of these plots are qualitatively
the same as in Fig.~\ref{plane_10}. Quantitatively, they differ
mainly in that the cosmologically favored regions are expanding
as $\tan\beta$ grows.
One reason is that the light stau becomes even
lighter as $\tan\beta$ increases, and this increases the neutralino
annihilation rate $\tz_1\tz_1\to \tau\bar{\tau}$ 
through $t$-channel stau exchange. In addition, the bottom and
tau Yukawa couplings increase with $\tan\beta$, which increases
the annihilation cross sections into $\tau$s and $b$s. 
Finally, the $H$ and $A$ Higgs boson masses are decreasing with $\tan\beta$,
and annihilation rates which proceed through these resonances increase. 
Co-annihilations again gives enhanced annihilation cross sections
on the left and farthest right hand sides of the allowed parameter space.
 The glitch in contours around $m_0\sim 2700$ GeV 
and $m_{1/2}\sim 425$ GeV occurs because $m_{\tz_1}\simeq m_t=175$ GeV,
so that $\sigma (\tz_1\tz_1\to t\bar{t})$ becomes large.
\begin{figure}[h]
\epsfxsize=8cm \epsfbox{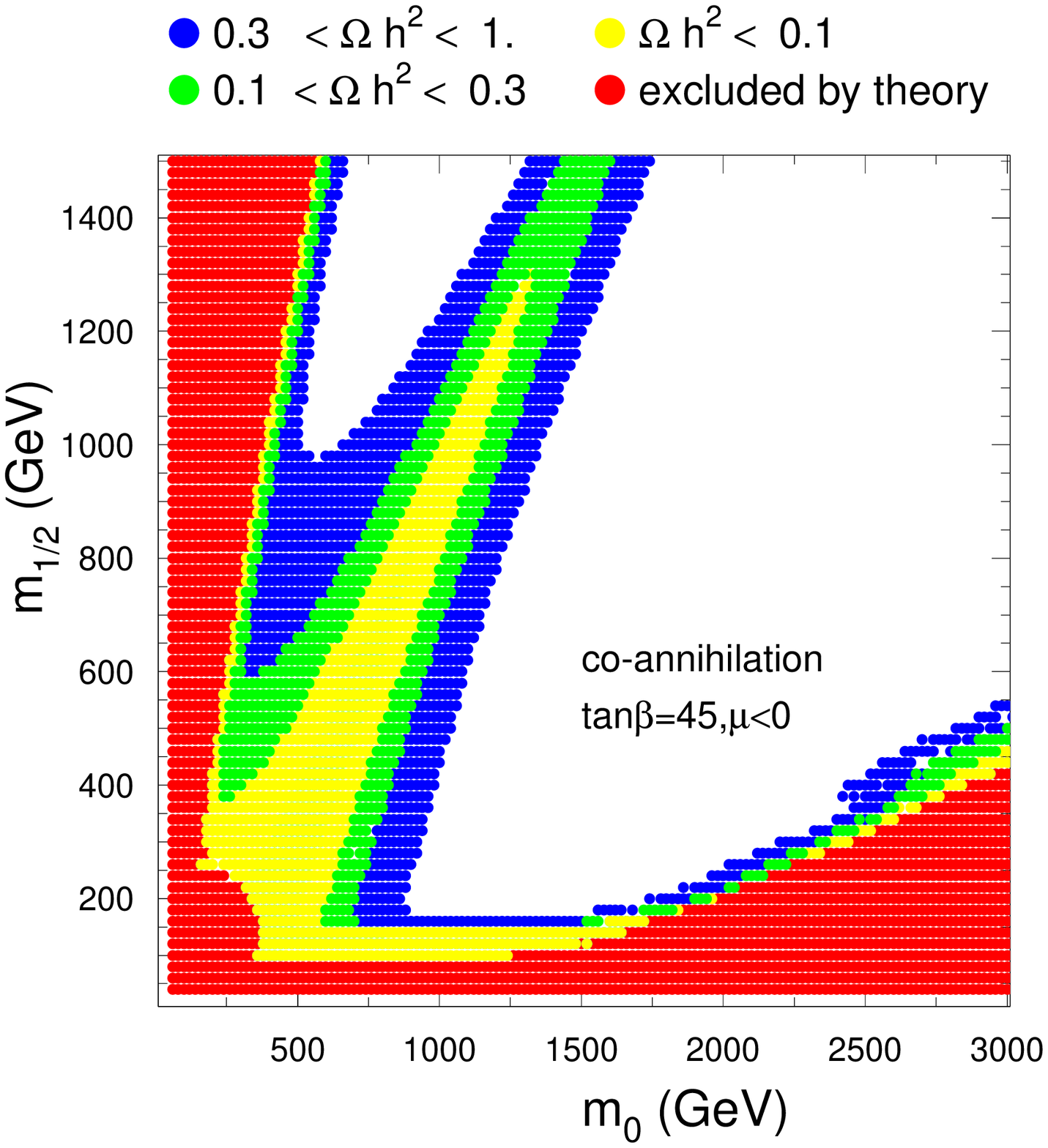}
\hspace*{-0.5cm}
\epsfxsize=8cm \epsfbox{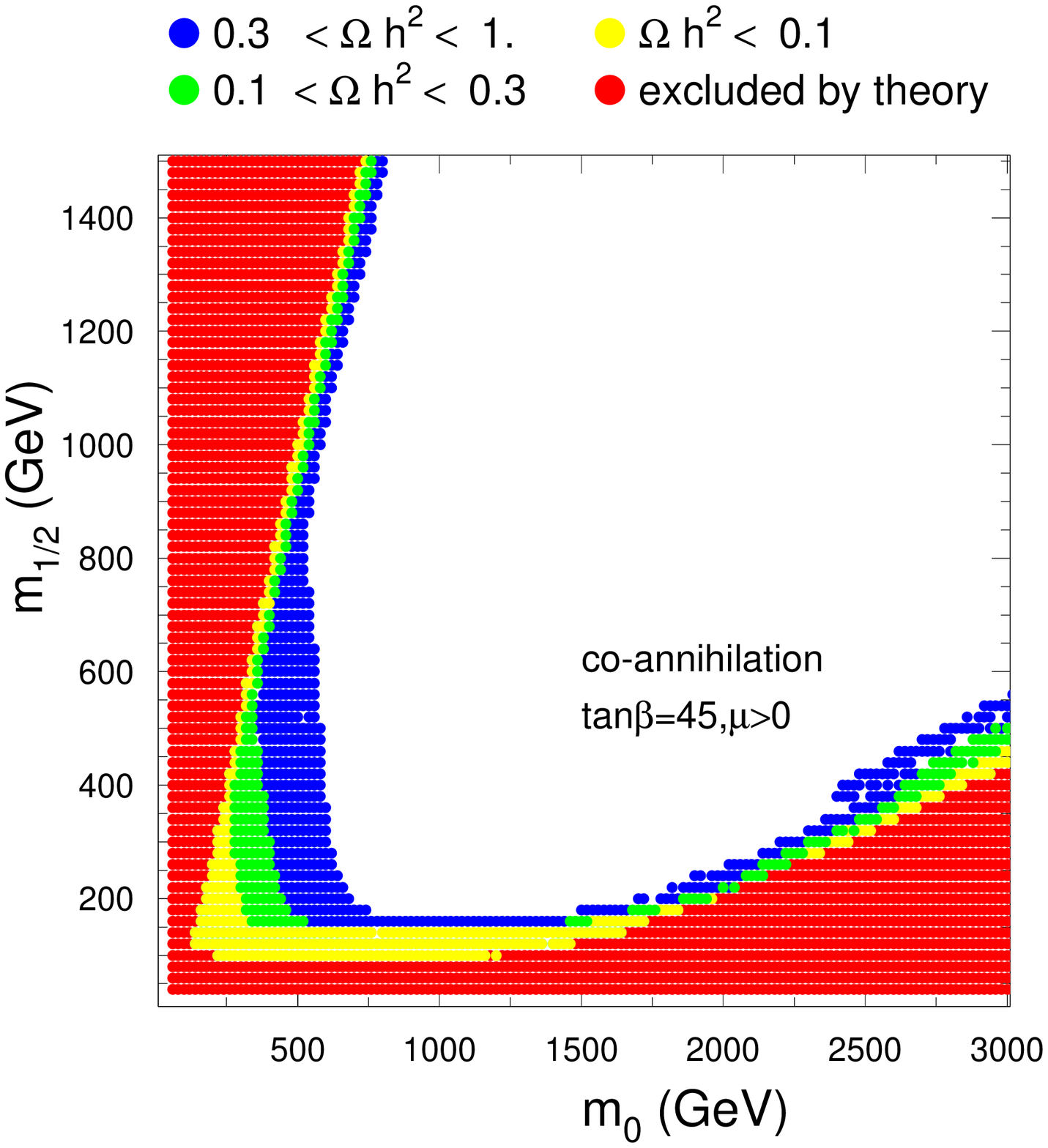}
\vskip -0.7cm
\caption{Regions of neutralino relic density in th
$m_0\ vs.\ m_{1/2}$ plane for $A_0=0$ and $\tan\beta =45$.}
\label{plane_45}
\end{figure}
\begin{figure}[h]
\vskip -0.7cm
\epsfig{file=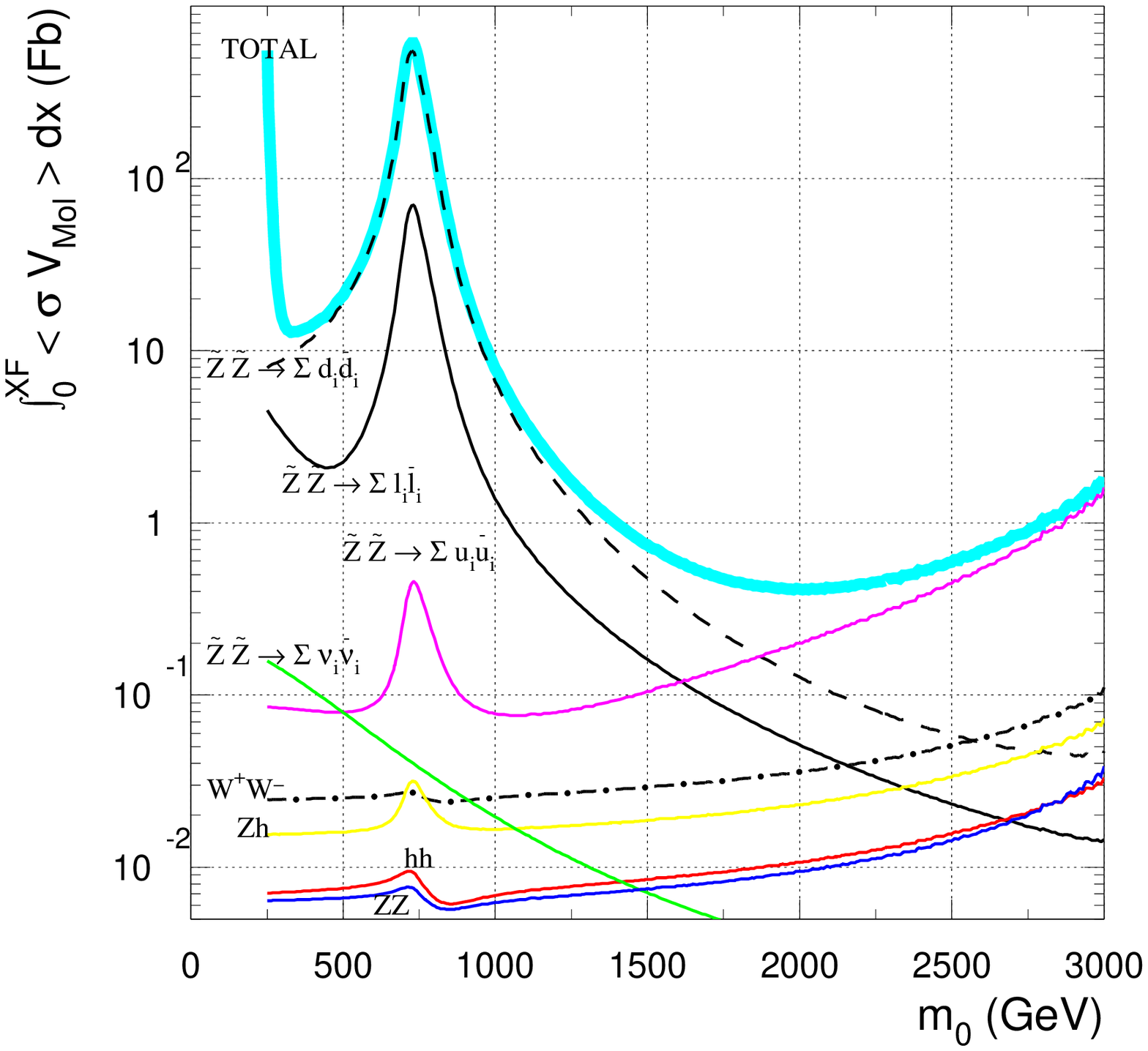,width=8.cm}
\hspace*{-0.5cm}
 \epsfig{file=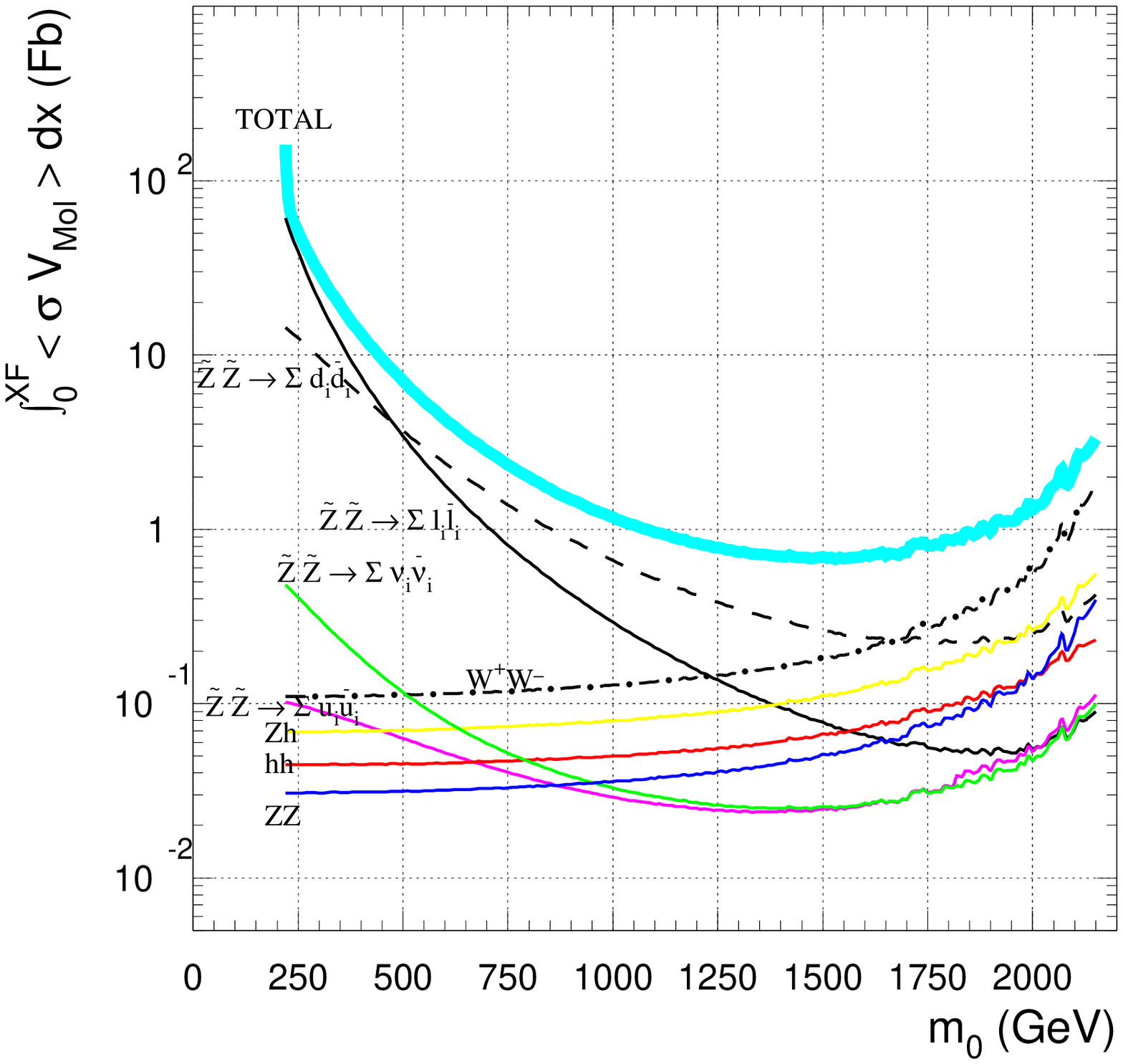,width=8.cm} 
\vskip -0.7cm
        \caption{Thermally averaged cross section times 
        velocity evaluated at $T_F$ for various 
        subprocesses. The thick light-grey(light-blue) curve  denotes the
        total of all annihilation and co-annihilation
        reactions. Left: 
        $m_{1/2}=600$ GeV, $\mu <0$, $A_0=0$ 
        and $\tan\beta =45$. Right: $m_{1/2}=300$ GeV, $\mu >0$, $A_0=0$ 
        and $\tan\beta =45$. }%
	\label{1d_sp}
\vskip -0.5cm
\end{figure}
In Fig.~\ref{plane_45}, we show the $m_0\ vs.\ m_{1/2}$ plane for
$\tan\beta =45$. In this case, the structure of the plane is changing
qualitatively, especially for $\mu <0$. First, there is a new
region of disallowed parameter space for $\mu <0$ in the lower left
due to $m_A^2<0$, which signals a breakdown of the REWSB mechanism.
Second, a corridor of very low relic density passes 
diagonally through the plot. The center of this region is where
$2m_{\tz_1}\simeq m_A$ and $m_H$. At the $A$ and $H$ resonance, there
is very efficient neutralino annihilation into $b\bar{b}$ final
states. This is illustrated in Fig.~\ref{1d_sp}(left), where we show the
integrated annihilation cross section times velocity versus $m_0$
for $m_{1/2}= 600$ GeV and $\mu <0$. At the very lowest values of $m_0$, there
is again the sharp peak due to neutralino-stau and stau-stau
co-annihilations.
 For larger values of $m_0$, however, 
the annihilation rate is dominantly into $b\bar{b}$ final states
over almost the entire $m_0$ range. This is due to the large
annihilation rates through the $s$-channel $A$ and $H$ diagrams,
even when the reactions occur off resonance. In this case, the widths
of the $A$ and $H$ are so large (both $\sim 10-40$ GeV across the range
in $m_0$ shown) that efficient $s$-channel annihilation can occur
throughout  considerable part of the parameter space, even when the resonance condition
is not exactly fulfilled. The resonance annihilation is explicitly 
displayed in this plot as the annihilation bump at $m_0$ just
below 1300 GeV. Another annihilation possibility is that
$\tz_1\tz_1\to b\bar{b}$ via $t$ and $u$ channel graphs.
In fact, these annihilation graphs are enhanced due to the large $b$
Yukawa coupling and decreasing value of $m_{\tb_1}$, but we have checked
that the $s$-channel annihilation is still far the dominant channel.
Annihilation into $\tau\bar{\tau}$ is the next most likely channel, but
is always below the level of annihilation into $b\bar{b}$ for the 
parameters shown in Fig.~\ref{1d_sp}(left). 
At even higher values of 
$m_0$ where the higgsino component of $\tz_1$ becomes non-negligible,
the annihilations into
$WW$ and $ZZ$ again become important; finally, at the highest values of
$m_0$, the $\tw_1$ and $\tz_2$ co-annihilation channels become large.

In Fig.~\ref{1d_sp}(right), we show again the subprocess annihilation rates
versus $m_0$ for $\tan\beta =45$, 
but this time for $\mu >0$ and for $m_{1/2}=300$ GeV.
 Although
no explicit resonance is evident for $\mu >0$, the dominant
annihilations are once again into $b\bar{b}$ final states over most of
the parameter space, due to the  wide Higgs resonances.
 To summarize the regions of mSUGRA model parameter space with 
reasonable values of neutralino relic density, we can
label four important regions: $i.$) annihilation through $t$-channel
slepton-- especially stau-- exchange, as occurs for low values
of $m_0$ and $m_{1/2}$, $ii.$) the stau co-annihilation region for
low values of $m_0$ on the edge of the excluded region, $iii.$) the large
$m_0$ region with non-negligible higgsino-component
annihilation, and also $\tw_1$
(and possibly $\tz_2$) co-annihilation occurs near the
edge of the limit of parameter space, and $iv.$) annihilation into
$b\bar{b}$ and $\tau\bar{\tau}$ final states through $s$-channel
$A$ and $H$ resonances at high $\tan\beta$.
 Other regions can include
top or bottom squark co-annihilation for large values of $A_0$, again
on the edge of parameter space where $\tst_1$ or $\tb_1$ become light,
or annihilation through $Z$ or $h$ resonances.
 The $Z-$resonance region
is essentially excluded now by constraints on sparticle masses from
LEP2.
\begin{figure}
\vskip -1cm
\epsfig{file=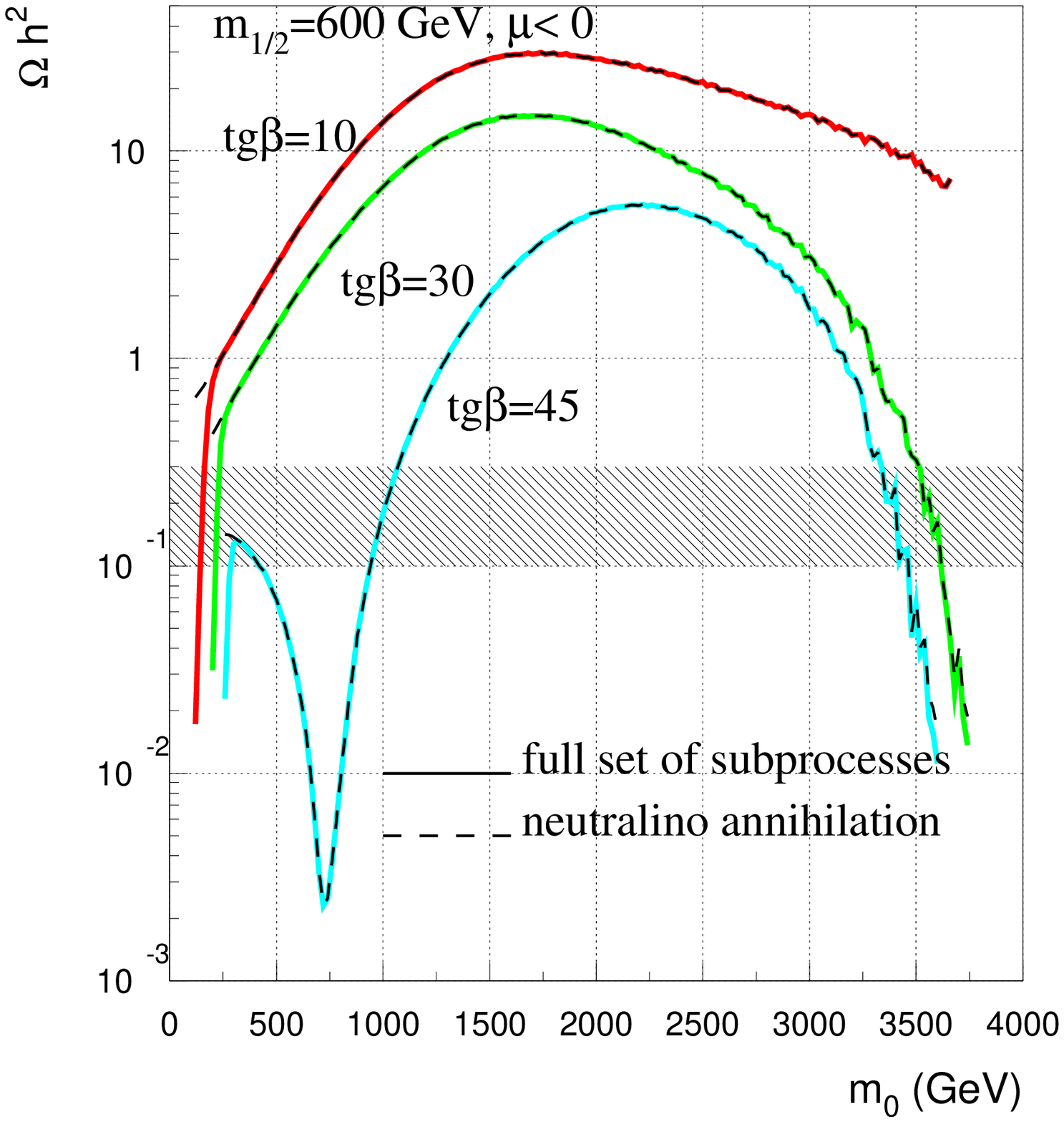,width=8cm,height=7.3cm} 
\hspace*{-0.5cm}
\epsfig{file=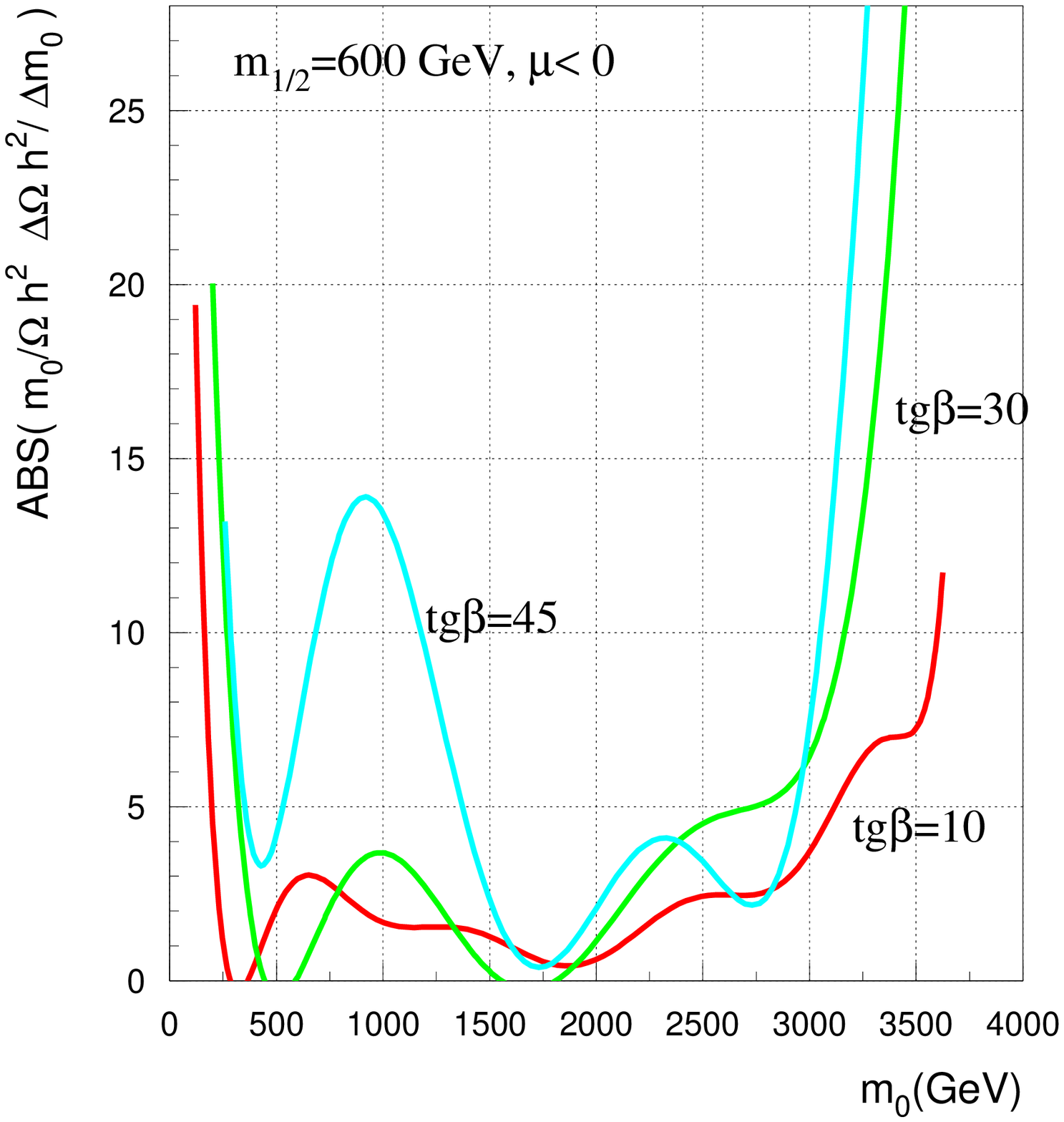,width=8cm,height=7.3cm}
\vskip -0.5cm
\caption{\label{1d_600} Neutralino relic density $\Omega_{\tz_1} h^2$(left) and 
the fine tuning parameter(right)[defined by Eq.(\ref{Eq:FineTuning})],
versus $m_0$ for $A_0=0$, $m_{1/2}=600$ GeV, $\mu <0$ 
and $\tan\beta =10$, 30 and 45.}%
\vskip -0.5cm
\end{figure}
\begin{figure}
\epsfig{file=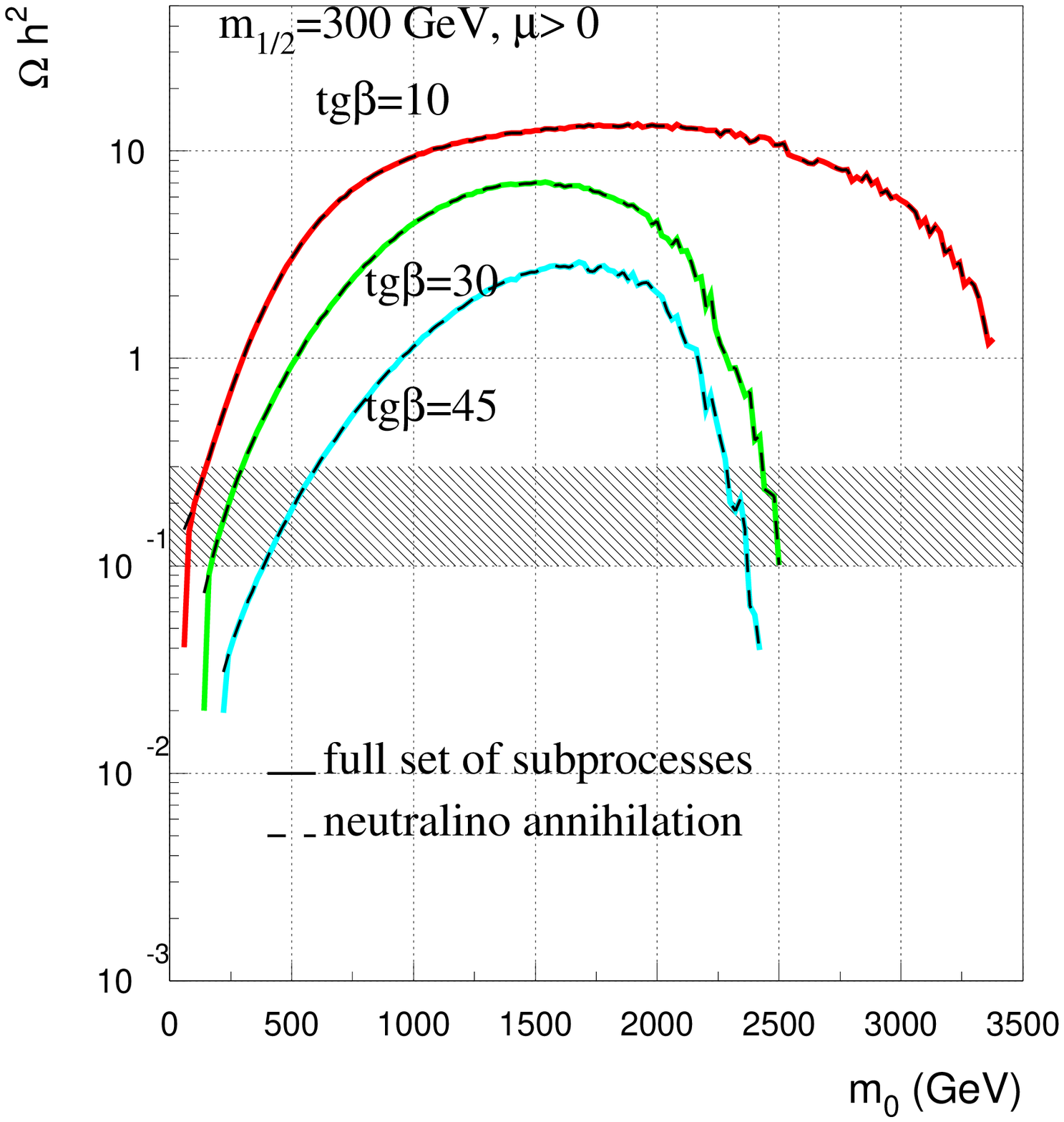,width=8cm,height=7.3cm} 
\hspace*{-0.5cm}
\epsfig{file=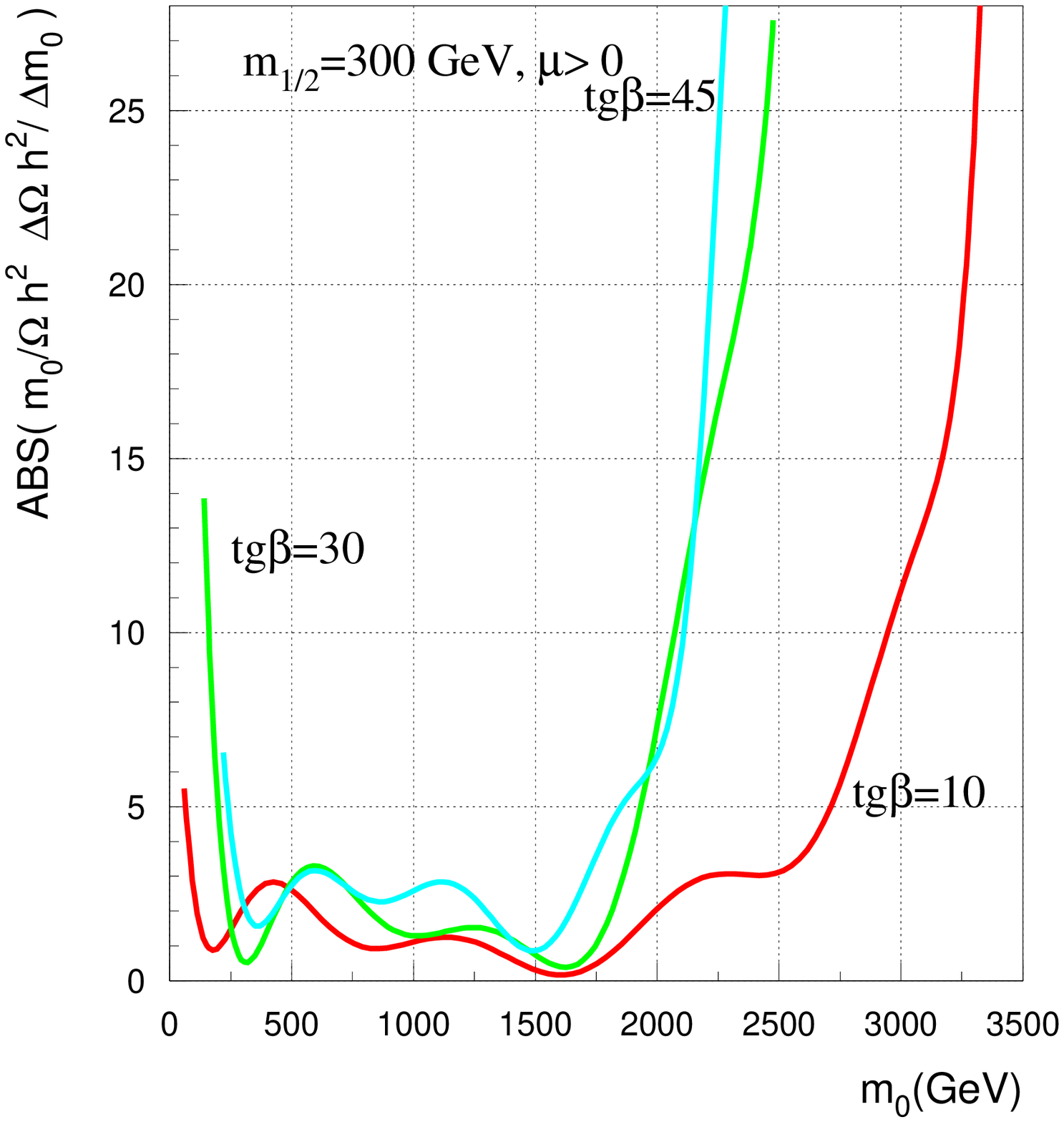,width=8cm,height=7.3cm}
\vskip -0.5cm
\caption{\label{1d_300}
The same as Fig.~\ref{1d_600} but for $m_{1/2}=600$ GeV and $\mu <0$.}%
\vskip -0.5cm
\end{figure}
It is useful to view the relic density $\Omega_{\tz_1}h^2$ directly as
a function of model parameters. We show in Fig.~\ref{1d_600}(left) the
value of $\Omega_{\tz_1}h^2$ versus the parameter $m_0$ for fixed
$m_{1/2}=600$ GeV, $A_0=0$, $\mu <0$ and for $\tan\beta =10$, 30 and 45.
The dashed curves show the result with no co-annihilations, while
the solid curves yield the complete calculation. The shaded band denotes the
cosmologically favored region with $0.1<\Omega_{\tz_1}h^2<0.3$. 
For this value of $m_{1/2}$, the lower $\tan\beta$ curves yield
a favored relic density only in the very low and very high $m_0$ regions,
and here the curves have a very sharp slope. The large slope is 
indicative of large fine-tuning, in that a small change of model parameters,
in this case $m_0$, yields a large change in $\Omega_{\tz_1} h^2$.
In contrast, the $\tan\beta =45$ curve shows a large region 
with good relic density and nearly zero slope($\mu>0$) or not very steep slope($\mu<0$),
and  hence with   little fine-tuning. 
In Fig.~\ref{1d_600}(right), we show the corresponding values of the fine-tuning,
basically the logarithmic derivative, as advocated by Ellis and 
Olive\cite{finetune}:
\begin{equation}
\Delta (m_0 )=\left|\frac{m_0}{\Omega_{\tz_1} h^2} 
\frac{\partial \Omega_{\tz_1} h^2}{\partial m_0}\right|
\label{Eq:FineTuning}
\end{equation}
As indicated earlier, the low fine-tuning regions mostly coincide with that of 
neutralino annihilation via $t$-channel slepton exchange (region $i.$)),
or off-resonance annihilation through $A$ and $H$ (region $iv.$)). 
The co-annihilation region $ii.)$ and focus point region $iii.$) tend to 
have higher fine-tunings due 
to the steep rise of the cross sections. Regions with simultaneous low fine-
tuning and preferred $\Omega_{\tz_1} h^2$ values are the 
best candidates for viable mSUGRA parameters.
In Fig.~\ref{1d_300}(left), we show $\Omega_{\tz_1}h^2$ versus $m_0$ for
$m_{1/2}=300$ GeV, $A_0=0$, $\mu >0$ and the same three $\tan\beta$
parameters. 
The curves reflect the broad regions of parameter space with reasonable
relic density values at high $\tan\beta$.
The corresponding plot of the fine-tuning parameter
is shown in Fig.~\ref{1d_300}(right). Again, there is large fine-tuning at the
edges of parameter space, but low fine-tuning in the intermediate
regions.
In conclusion, the relic density and the fine-tuning parameter together 
tend to prefer mSUGRA model parameters in regions $i.$) or $iv.$).
These two regions lead to distinct collider signatures for future
searches for supersymmetric matter.
{ \ \vspace*{-0.5cm}}
\section{Conclusions}

In conclusion, we have performed a calculation of the neutralino
relic density in the minimal supergravity model including all
$2\to 2$ neutralino annihilation and co-annihilation processes, where
the initial state includes $\tz_1$, $\tz_2$, $\tw_1$, $\te_1$, $\tmu_1$,
$\ttau_1$, $\tst_1$ and $\tb_1$. 
The calculation was performed using the
CompHEP program for automatic evaluation of Feynman diagrams, coupled
with ISAJET for sparticle mass evaluation in the mSUGRA model, and
for standard and supersymmetric couplings and decay widths. 
We implemented relativistic thermal averaging,
which is especially important for evaluating the relic density
when resonances in the annihilation cross section are present, and
neutralino thermal velocities can be relativistic. The three-dimensional
integration was
performed by Monte Carlo evaluation with importance sampling, which
yields in general good convergence even in the presence of
narrow resonances. We note that a calculation of
similar scope and procedure was recently reported in Ref. \cite{belanger}.

We found four regions of parameter space that led to relic densities
in accord with results from cosmological measurements, {\it i.e.}
$0.1<\Omega_{\tz_1}h^2<0.3$. These include {\it i}.) the region
dominated by $t$-channel slepton exchange, {\it ii.}) the region
dominated by stau co-annihilation, {\it iii.}) the large $m_0$ region
dominated by a more higgsino-like neutralino and {\it iv.}) the
broad regions at high $\tan\beta$ dominated by off-shell 
annihilation through the $A$ and $H$ Higgs boson resonances.
Regions {\it ii.}) and {\it iii.}) generally have large
fine-tuning associated with them, and although it is logically
possible that nature has chosen such parameters, any slight deviation
of model parameters would lead to either too low or too high a relic
density. Region {\it i.}) generally has the property that some of the
sleptons have masses less than about 300-400 GeV. This region can give 
rise to a rich set of collider signatures, since many of the sparticles
are relatively light. 

Region {\it iv.}) gives broad regions of 
model parameter space with reasonable values of relic density as well
as low values of the fine-tuning parameter. 
It can also allow quite heavy values of SUSY particle masses, 
which would be useful to suppress many flavor-violating 
(such as $b\to s\gamma$)\cite{bsg} and CP violating
loop processes, and the muon $g-2$ value\cite{gm2}. 
In many respects region {\it iv.}) is a favored region of parameter
space. The neutralino relic density may well point the way to the sort of
SUSY signatures we should expect at high energy collider experiments.

\bigskip
We thank Manuel Drees, Konstantin Matchev, Leszek Roszkowski 
and Xerxes Tata for discussions.
This research was supported in part by the U.S. Department of Energy
under contract number DE-FG02-97ER41022.


\begin{thebibliography}{999}

\bibitem{cmb} 
A.~T.~Lee {\it et al.} (MAXIMA Collaboration),
Astrophys.\ J.\  {\bf 561}, L1 (2001);
C. B. Netterfield {\it et al.} (BOOMERANG Collaboration), 
astro-ph/0104460 (2001); N. W. Halverson {\it et al.} (DASI Collaboration),
astro-ph/0104489 (2001); P. de Bernardis {\it et al.}, astro-ph/0105296 
(2001).
%
\bibitem{wlf} See {\it e.g.} W. L. Freedman, Phys. Rept. {\bf 333}, 13 (2000).
%
\bibitem{supernovae} A. G. Riess {\it et al.}, Astron. J. {\bf 116}, 
1009 (1998); S. Perlmutter {\it et al.}, Astrophys. J. {\bf 517}, 565 (1999).
%
\bibitem{jaffe} A. H. Jaffe {\it et al.}, Phys. Rev. Lett. {\bf 86}, 3475 
(2001).
%
\bibitem{nucleo} K. Olive, G. Steigman and T. Walker,
Phys. Rept {\bf 333-334}, 389 (2000); S. Burles, K. Nollet
and M. Turner, Phys. Rev. Lett. {\bf 82}, 4176 (1999) and
Phys. Rev. D{\bf 63}, 063512 (2001); 
D. Tytler, J. O'Meara,  N. Suzuki and D. Lubin,
astro-ph/0001318 (2000).
%
\bibitem{review} For a review, see {\it e.g.} M. S. Turner,
astro-ph/0108103 (2001).
%
\bibitem{jkg} For a review, see G. Jungman, M. Kamionkowski and K. Griest,
Phys. Rept. {\bf 267}, 195 (1996).
%
\bibitem{sugra} A. Chamseddine, R. Arnowitt and P. Nath,
Phys. Rev. Lett. {\bf 49}, 970 (1982);
R. Barbieri, S. Ferrara and C. Savoy, Phys. Lett. {\bf B119}, 343 (1982);
L.J. Hall, J. Lykken and S. Weinberg, Phys. Rev. {\bf D27}, 2359 (1983).
%
\bibitem{isajet} H. Baer, F. Paige, S. Protopopescu and X. Tata,
hep-ph/0001086 (2000).
%
\bibitem{early} H. Goldberg, Phys. Rev. Lett. {\bf 50}, 1419 (1983);
J. Ellis, J. Hagelin, D. Nanopoulos and M. Srednicki, 
Phys. Lett. {\bf B127}, 233 (1983);
J. Ellis, J. Hagelin, D. Nanopoulos, K. Olive and M. Srednicki, 
Nucl. Phys. {\bf B238}, 453 (1984).
%
\bibitem{ows} M. Srednicki, R. Watkins and K. Olive,
Nucl. Phys. {\bf B310}, 693 (1988).
%
\bibitem{barb} 
R. Barbieri, M. Frigeni and G. F. Giudice, Nucl. Phys. {\bf B313}, 725 (1989);
%
\bibitem{gkt}
K. Griest, M. Kamionkowski and M. Turner, Phys. Rev. {\bf D41}, 3565 (1990).
%
\bibitem{gs} K. Griest and D. Seckel, Phys. Rev. D{\bf 43}, 3191 (1991).
%
\bibitem{gg} P. Gondolo and G. Gelmini, Nucl. Phys. {\bf B360}, 145 (1991).
%
\bibitem{bottino} A. Bottino, V. de Alfaro, N. Fornengo, G. Mignola 
and S. Scopel, Astropart. Phys. {\bf 1}, 61 (1992); A. Bottino {\it et al.},
Astropart. Phys. {\bf 2}, 67 (1994); V. Berezinsky {\it et al.}, 
Astropart. Phys. {\bf 5}, 1 (1996).
%
\bibitem{dn} M. Drees and M. Nojiri, Phys. Rev. {\bf D47}, 376 (1993).
%
\bibitem{leszek} J. Ellis and L. Roszkowski, 
Phys. Lett. {\bf B283}, 252 (1992); L. Roszkowski and R. Roberts, 
Phys. Lett. {\bf B309}, 329 (1993); G. Kane, C. Kolda, 
L. Roszkowski and J. Wells, Phys. Rev. D{\bf 49}, 6173 (1994).
%
\bibitem{an} P. Nath and R. Arnowitt, Phys. Rev. Lett. {\bf 70}, 3696 (1993);
R. Arnowitt and P. Nath, Phys. Lett. {\bf B437}, 344 (1998).
%
\bibitem{bb} H. Baer and M. Brhlik, Phys. Rev. D{\bf 53}, 597 (1996) 
and Phys. Rev. D{\bf 57}, 567 (1998); 
H.~Baer, M.~Brhlik, M.~A.~Diaz, J.~Ferrandis, 
P.~Mercadante, P.~Quintana and X.~Tata,
Phys.\ Rev.\ D {\bf 63}, 015007 (2001)
%
\bibitem{eg} J. Edsj\"o and P. Gondolo, Phys. Rev. D{\bf 56}, 1879 (1997). 
%
\bibitem{bk} V. Barger and C. Kao, Phys. Rev. D{\bf 57}, 3131 (1998) and
Phys. Lett. {\bf B518}, 117 (2001).
%
\bibitem{ellis} J. Ellis, T. Falk, G. Ganis, K. Olive and M. Srednicki,
Phys. Lett. {\bf B510}, 236 (2001).
%
\bibitem{darksusy} DarkSUSY, by P. Gondolo and J. Edsj\"o,
astro-ph/0012234 (2000).
%
\bibitem{fmw} J. Feng, K. Matchev and F. Wilczek, 
Phys. Lett. {\bf B482}, 388 (2000) and Phys. Rev. D{\bf 63}, 045024 (2001);
see also hep-ph/0111295 (2001).
%
\bibitem{ellis_co} J. Ellis, T. Falk and K. Olive, Phys. Lett. {\bf B444},
367 (1998); J. Ellis, T. Falk, K. Olive and M. Srednicki, 
Astropart. Phys. {\bf 13}, 181 (2000). 
%
\bibitem{an2} R. Arnowitt, B. Dutta and Y. Santoso, Nucl. Phys. {\bf B606}, 
59 (2001). 
%
\bibitem{pallis} M. Gomez, G. Lazarides and C. Pallis, 
Phys. Rev. D{\bf 61}, 123512 (2000) and Phys. Lett. {\bf B487}, 313 (2000).
%
\bibitem{leszek2} L. Roszkowski, R. Ruiz de Austri and T. Nihei,
JHEP{\bf 0108}, 024 (2001).
%
\bibitem{manuel} A. Djouadi, M. Drees and J. Kneur, 
JHEP{\bf 0108}, 055 (2001).
%
\bibitem{drees} C. Boehm, A. Djouadi and M. Drees, Phys. Rev. {\bf D62}, 
035012 (2000).
%
\bibitem{santoso} J. Ellis, K. Olive and Y. Santoso, hep-ph/0112113 (2001).
%
\bibitem{comphep} CompHEP~v.33.23, by A. Pukhov {\it et al.}, hep-ph/9908288 (1999).
%
\bibitem{belanger} G. Belanger, F. Boudjema, A. Pukhov and A. Semenov,
hep-ph/0112278 (2001).
%
\bibitem{roszkowski} T. Nihei, L. Roszkowski and R. R. de Austri, 
hep-ph/0202009 (2002).
%
\bibitem{bases} S.~Kawabata,
{\it Prepared for 2nd International Workshop on Software Engineering, 
Artificial Intelligence and Expert Systems for High-energy and Nuclear
Physics, La Londe Les Maures, France, 13-18 Jan 1992}.
%
\bibitem{finetune} J. Ellis and K. Olive, Phys. Lett. {\bf B514}, 114 (2001).
%
\bibitem{lep2} See {\it e.g.} R. Barate {\it et al.}, 
Phys. Lett. {\bf B499}, 67 (2001).
%
\bibitem{lep2higgs} For combined LEP2 limits on MSSM Higgs bosons, 
see hep-ex/0107030 (2001).
%
\bibitem{tevatron} H. Baer, M. Drees, F. Paige, P. Quintana and X. Tata,
Phys. Rev. D{\bf 61}, 095007 (2000);
V. Barger and C. Kao, Phys. Rev. D{\bf 60}, 115015 (1999);
K. Matchev and D. Pierce, Phys. Lett. {\bf B467}, 225 (1999); 
for a review, see S. Abel {\it et al.},
hep-ph/0003154 (2000).
%
\bibitem{feng} J. Feng, K. Matchev and T. Moroi, Phys. Rev. D{\bf 61}, 
075005 (2000).
%
\bibitem{lhc} H. Baer, C. H. Chen, F. Paige and X. Tata,
Phys. Rev. D{\bf 52}, 2746 (1995) and Phys. Rev. D{\bf 53}, 6241 (1996);
H. Baer, C. H. Chen, M. Drees, F. Paige and X. Tata,
Phys. Rev. D{\bf 59}, 055014 (1999).
%
\bibitem{bmt} H. Baer, R. Munroe and X. Tata, Phys. Rev. D{\bf 54}, 
6735 (1996); Erratum Phys. Rev. D{\bf 56}, 4424 (1997). 
%
\bibitem{bsg} See {\it e.g.} H. Baer, M. Brhlik, D. Castano and
X. Tata, Phys. Rev. D{\bf 58}, 015007 (1998).
%
\bibitem{gm2} See {\it e.g.} H. Baer, C. Balazs, J. Ferrandis and
X. Tata, Phys. Rev. D{\bf 64}, 035004 (2001).
%

\end{thebibliography}
\end{document}